\documentclass{article}
\usepackage{amsmath}
\def\p{\partial}

\def\G{\Gamma}
\def\g{\gamma}

\def\ld{\lambda}
\def\L{\Lambda}

\def\s{\sigma}

\def\b{\beta}

\def\a{\alpha}

\def\pdellx'{\frac{\partial}{\partial x'}}
\def\pdellw'{\frac{\partial}{\partial w'}}

\newcommand{\be}{\begin{equation}}
\newcommand{\ee}{\end{equation}}
\def\bed{\begin{displaymath}}
\def\eed{\end{displaymath}}
\def\bea{\begin{eqnarray}}
\def\eea{\end{eqncrray}}
\def\[{$$}
\def\]{$$}
\newcommand{\beas}{\begin{eqnarray*}}
\newcommand{\eeas}{\end{eqnarray*}}

\newcounter{stokes}

\begin{document}
\title{ YANG-MILLS GRAVITY IN FLAT SPACE-TIME, I.  \\
 CLASSICAL GRAVITY WITH TRANSLATION GAUGE SYMMETRY }

\author{ Jong-Ping Hsu\footnote{E-mail: jhsu@umassd.edu} \\
 Department of Physics,
 University of Massachusetts Dartmouth \\
 North Dartmouth, MA 02747-2300, USA }


\maketitle
{\small We formulate and explore the physical implications of a new translation 
gauge theory of gravity in flat space-time with a new Yang-Mills 
action, which involves quadratic gauge curvature and fermions.  The 
theory shows that the presence 
of an "effective Riemann metric tensor" for the motions of classical 
particles and light rays is probably the manifestation of the translation gauge 
symmetry.  In the post-Newtonian approximation of the tensor gauge 
field produced by the energy-momentum tensor, the results are shown to 
be consistent with classical tests of gravity and with the quadrupole 
radiations of binary pulsars.}
 
\bigskip
 

\bigskip


 
\section{ Introduction}

 Since the creation of the Yang-Mills theory in 1954, quantum field theory 
with gauge symmetry based on a flat space-time have been successfully applied to all 
fundamental physical theories of fields, except for the gravitational field. 
Dyson stressed that the most glaring incompatibility of concepts in 
contemporary physics is that between the principle of general coordinate 
invariance and a quantum-mechanical description of all of nature.~\cite{1} 
Quantum gravity appears to be the last challenge to the powerful gauge symmetry of 
the Yang-Mills theory.

In contrast to the field-theoretic approach, Einstein's approach involving the geometrization 
of physics appears to be successful only in the formulation of classical gravity. 
If the electromagnetic force, which is velocity-dependent, is treated 
in the same way, it 
is natural to employ the Finsler geometry rather than the Riemannian 
geometry.~\cite{2,3,4} The fundamental metric tensors of the Finsler geometry depend on both position and 
velocity (i.e., the differential of the coordinates). So far, however, all attempts 
to geometrize classical electrodynamics have been unsuccessful in the sense that they have not 
even begun to approach the usefulness of quantum electrodynamics. Thus, it is interesting to 
investigate the possibility of understanding gravity within the framework of 
the usual physical field theory, especially a framework with gauge symmetry in 
a flat space-time. An obvious challenge is to construct a gauge-invariant action 
(which involves quadratic gauge-curvature) in flat space-time that can produce 
the same good agreement with experimental results as those in general 
relativity (whose action involves   
linear space-time curvature~\cite{5,6,7,8,9,10,11}).  

In this paper, we do just that. We generalize the usual Yang-Mills theory with internal 
gauge groups to a gauge theory with the external space-time translation group 
in which the generators of 
the group do not have constant matrix representations. This generalization appears 
to be essential for tensor gauge  fields to be generated by the 
well-defined and conserved energy-momentum tensor.  We investigate a physical system 
involving fermion matter which generates a gravitational tensor field.  We find an interesting action with 
a quadratic gauge-curvature in flat space-time and with space-time 
translation gauge symmetry. This action leads to a linearized gauge field equation 
that is exactly the same as the linearized Einstein equation, and is in agreement with 
experimental tests of general relativity such as the red shift. To at least the 
second order approximation, Yang-Mills gravity also 
predicts correctly the perihelion 
shift of Mercury and the gravitational quadrupole radiation of binary pulsars. 
This theory of Yang-Mills gravity with translation gauge symmetry is interesting 
because the gauge symmetry in flat space-time appears to be crucial for a quantum 
field theory to be unitary and renormalizable.

\section{Translation Gauge Transformations and Accelerated Frames}

The formulations of
electromagnetic and Yang-Mills theories associated with internal gauge
groups are all based on the replacement,
$\p_\mu \to \p_\mu+igB_\mu$.  The field $B_\mu=B_\mu^a t_a$
involves constant matrix representations of
the generators $t_a$ of the gauge groups.  However, the generators of the
external space-time translation group T(4) are the displacement operators, $p^\mu=i\p ^\mu$ \
(c=$\hbar$=1); thus, the replacement takes a different form,
\be
 \p ^\mu
\to \p ^\mu  -ig\phi^{\mu\nu} p_\nu \equiv J^{\mu\nu}\p_\nu, \ \ \ \  
J^{\mu\nu}=\eta^{\mu\nu}+g\phi^{\mu\nu},
\ee
in such a gauge theory in inertial frames, where $\eta^{\mu\nu}=(1,-1,-1,-1)$.
Since the generators of this external translation group T(4) is $p^\mu=i\p^\mu$, 
we have a symmetric tensor gauge field 
$\phi^{\mu\nu}=\phi^{\nu\mu}$ (i.e., a spin-2 field) rather than a 4-vector 
field (i.e., a 
spin-1 field) in the gauge covariant derivative, $\Delta^\mu = 
J^{\mu\nu} \p_\nu$ (in inertial frames).  Nevertheless, we follow the Yang-Mills approach to
formulate a theory of gravity based on T(4) in flat space-time.  It is
 precisely this unique property (1) due to 
displacement operators of the T(4) group that naturally leads to an effective Riemannian metric 
tensor,\footnote{The speculation that Einstein's theory of gravity 
may be an effective field theory has been around for decades among  
theorists.  The idea of an 
effective Riemannian space due to the
presence of the gravitational field in Minkowski space-time was extensively discussed by
Logunov, Mestvirishvili and others.  Their theory
is not based on the space-time translation
gauge group.  Moreover,  they postulated a different gravitational action, which involves a linear
`scalar curvature of effective Riemannian space.'  See, for example, 
Ref. 12. } and a universal repulsive force for all matter and 
anti-matter, where the force is characterized by a coupling constant 
$g$ with the dimension of length in natural units.  This is in sharp 
contrast with all 
coupling constants, which are dimensionless, in the usual Yang-Mills theories.     For external gauge groups related to space-time, e.g.,
the de Sitter group or the Poincar\'{e} group, the gauge invariant Lagrangian
involving fermions turns out to be
richer in content.\footnote{For example, in order for the 
fermion Lagrangian to be de Sitter gauge invariant,
apart from the presence of the usual Yang-Mills (`phase') fields, there must also beÊ
distinct ``scale fields'' to compensate the non-commutativity between
Dirac's matrices and the generators of the de Sitter
groups.  It suggests the existence of an additional gravitational spin
force.  See Ref. 13. }  If the space-time translation group T(4) is replaced by 
the Poincar\'{e} group, we expect that a new gravitational spin force 
generated by  the fermion spin density 
will appear in the theory.~\cite{13}  However, in this paper, we concentrate on
the external gauge group of space-time translations T(4), which
is the Abelian subgroup of the Poincar\'{e} group and is non-compact. 
This group T(4) is particularly interesting because it is the
minimal group related to the conserved energy-momentum tensor, which
couples to a tensor (or
spin-2) field $\phi_{\mu\nu}$ and is the source of gravity.

Translation gauge symmetry is based on the local space-time
 translation with an arbitrary infinitesimal
vector gauge-function $\L^\mu (x)$,
\be
x^\mu \to x'^\mu = x^\mu + \L^\mu(x), \ \ \ \ \  x \equiv
x^\lambda=(w,x,y,z).
\ee
It is interesting that this transformation
 has a dual interpretation: (i) a shift of the space-time
coordinates by an infinitesimal vector gauge-function $\Lambda^\mu (x)$, and
(ii) an arbitrary infinitesimal transformation.  For the 
interpretation (ii) to be consistent in the theory, we must formulate 
Yang-Mills gravity for both inertial and non-inertial frames (i.e.,  
general frames).  Fortunately, we can accommodate
these two mathematical implications of the transformation (2) by defining a
gauge transformation of space-time translations for physical quantities
$Q^{\mu_1 ...\mu_m}_{\a_1 ...\a_n} (x)$ in the
 Lagrangian of fields:
$$
Q^{\mu_1 ...\mu_m}_{\a_1 ...\a_n} (x) \to \left(Q^{\mu_1 ...\mu_m}_{\a_1
...\a_n} (x)\right)^{\$} $$
\be
= \left(Q^{\nu_1 ...\nu_m}_{\b_1 ...\b_n} (x) - \L^\lambda (x)
\p_\ld Q^{\nu_1 ...\nu_m}_{\b_1 ...\b_n} (x)\right)
\frac{\p x'^{\mu_1}}{\p
x^{\nu_1}}... \frac{\p
x'^{\mu_m}}{\p x^{\nu_m}}
\frac{\p x^{\b_1}}{\p x'^{\a_1}} ... \frac{\p x^{\b_n}}{\p
x'^{\a_n}},
\ee
where $\mu_1, \nu_1, \a_1,\b_1,$ etc. are space-time indices.  As 
usual, both (Lorentz) spinor field $\psi$ and (Lorentz) scalar field 
$\Phi$ are treated as `coordinate scalars' and have  the 
same translational gauge transformation:
$\psi \to \psi^{\$}=\psi - \L^\ld \p_\ld \psi, \ \ \  \Phi \to 
\Phi^{\$}=\Phi - \L^\ld \p_\ld \Phi.$  In general, the gauge transformations for scalar, vector and tensor 
fields are given by
$$
Q(x) \to (Q(x))^{\$}=Q(x) - \L^\ld \p_\ld Q(x), \ \ \ \ \  Q(x)=\psi, 
\overline{\psi}, \Phi,
$$
$$D_\mu Q \to (D_\mu Q)^{\$}=D_\mu Q - \L^\ld \p_\ld (D_\mu
Q) - (D_\ld Q) \p_\mu \L^\ld,
$$
\be
\G^\mu \to (\G^\mu)^{\$} =\G^\mu - \L^\ld \p_\ld \G^\mu + \G^\ld \p_\ld \L^\mu,
\ee
$$T_{\mu\nu} \to (T_{\mu\nu})^{\$}=T_{\mu\nu} - \L^\ld \p_\ld T_{\mu\nu} -
T_{\mu\a} \p_\nu
\L^\a - T_{\a\nu} \p_\mu \L^\a, \ \ \ \  T_{\mu\nu}= J_{\mu\nu}, \
P_{\mu\nu},
$$
$$Q^{\mu\nu} \to (Q^{\mu\nu})^{\$}= Q^{\mu\nu} - \L^\ld \p_\ld Q^{\mu\nu} +
Q^{\ld\nu} \p_\ld \L^\mu + Q^{\mu\ld} \p_\ld \L^\nu.
$$
Here $ D_\mu$ denotes the partial covariant derivative 
associated
with a metric tensor $P_{\mu\nu} (x)$ in a general reference frame
(inertial or non-inertial).  Note that the functions $D_{\mu}Q$ and
$D_{\mu} D_{\nu}Q$ transform, by definition, as a  
covariant vector $Q_\mu (x)$ and a covariant tensor $Q_{\mu\nu}(x)$ 
respectively under the translational gauge
transformation.   The change of variables in the translation
gauge transformation (4) in flat space-time is formally similar to the Lie variations in the
coordinate transformations in Riemannian geometry.

Since the theory of Yang-Mills gravity should be formulated in a general
frame of reference (inertial or non-inertial) characterized by a certain
metric tensor $P_{\mu\nu}$, let us consider specific examples
of $P_{\mu\nu} (x)$ for
general frames.  To substantiate  the existence of such a metric 
tensor $P_{\mu\nu}$, let us
consider a general-linear-acceleration transformation between an inertial
frame
$F_I(w_I,x_I,y_I,z_I)$ and a non-inertial frame $F(w,x,y,z)$.  

Suppose
$F_I$ is at rest and the frame $F$ moves in
the x-direction with an initial velocity $\b_o$ and an arbitrary
linear acceleration $\a(w)$.
The general-linear-acceleration transformations~\cite{14,15} between an inertial frame $F_I$ and a general 
frame  $F$ (which moves with an arbitrary
velocity $\b (w)$ in the +x-direction) are given by 
\be
w_I = \g \b(w)U
- \frac{\b_o}{\a_o \g_o}, \ \ x_I=\g U - \frac{1}{\a_o \g_o}, \ \ y_I= y, \ \ z_I = 
z; 
\ee
$$
\g=\frac{1}{\sqrt{1-\b^2(w)}}, \ \ \ \g_{o}=\frac{1}{\sqrt{1-\b_{o}^2}}, \ \ 
\   U=x+ \frac{1}{\a (w) \g_o^2}, \ \ \  \a (w) =\frac{d\b (w)}{dw}. 
$$
They reduce to the Wu transformation in the limit of 
constant-linear-acceleration, i.e., $\a(w) \to \a_{o}$,~\cite{16,17,18}
$$
w_I = \g (\b_{o} + 
\a_{o}w)(x+\frac{1}{\a_{o}\g_{o}^{2}})-\frac{\b_{o}}{\a_o \g_o},$$
\be
x_I=\g (x+\frac{1}{\a_{o}\g_{o}^{2}}) - \frac{1}{\a_o \g_o}, \ \ y_I= y, \ \ z_I = 
z; 
\ee
In the special case $\b_{o}=0$, the Wu transformation (6) becomes the M$\o$ller 
transformation,~\cite{19,20,21,22}
$$
w_I = (x+\frac{1}{\a_{o}})sinh(\a_{o}w^{*}), \ \ \ \
x_I= (x+\frac{1}{\a_{o}})cosh(\a_{o}w^{*})- \frac{1}{\a_o}, \ \ y_I= y,
\ \ z_I = z, 
$$
provided one makes a change of time variable,
$w=(1/\a_{o}) tanh(\a_{o} w^{*})$.  Furthermore, 
the Wu transformation (6) reduces to the 
Lorentz transformation in the limit of zero acceleration, $\a_{o} 
\to  0$,
\be
w_I = \g_{o}(w +\b_{o}x),
\ \ x_I=\g_{o}(x+\b_{o} w), \ \ y_I= y, \ \ z_I = 
z; 
\ee 
One can verify that the general-linear-acceleration transformation 
(5) preserves the space-time
interval $ds^2$:
\be
 ds^2=dw_I{^2} -dr_I{^2}= W^2 dw^2 + 2U_{J}dwdx - dx^2 - dy^2 - dz^2 =
 P_{\mu\nu}dx^{\mu}dx^{\nu},
\ee
$$P_{00}=W^{2}, \ \ P_{01}=P_{10}=U_{J}, \ \  P_{11}=P_{22}=P_{33}=-1,$$
$$
W^2= W_c^2 - U_{J}^2,  \ \ \ \ 
W_c = \g^2 \left(\frac{1}{\g_o^2} +\a (w)
 x\right), \ \ \ \ \
U_{J}=\frac{d\a/dw}{\a^2(w)\g^2_o},
$$
where $dx^{\mu}=(dw,dx,dy,dz)$ and the non-vanishing components of 
$P_{\mu\nu}$ are $P_{00}=W^{2}, P_{01}=P_{10}=U_{J},$ etc.  for 
general-linear-acceleration frames.  It is interesting to see that 
the metric tensors $P_{01}$ involves  the `jerk', i.e., the time derivative 
of the acceleration $d\a(w)/dw$, which appears very rarely in physics.  
Moreover, equations (6) and (8) indicate that all constant-linear-acceleration frames
have the metric tensor of the form
$P_{\mu\nu}=(W_c^2,-1,-1,-1)$ because $d\a(w)/dw=0$ and $\a(w)=\a_{o}¥$.~\cite{18}
The existence of the finite space-time transformations between  inertial
frames and general-linear-acceleration frames 
 implies that the space-time associated  with the
general-linear-acceleration (and constant-linear-acceleration) frames 
is flat.~\cite{20}  Thus, the physical space-time of all  these general frames 
is characterized by
 zero Riemann-Christoffel curvature tensor, 
 $R^{\ld}_{\mu\nu\a}=0$.    

It turns out that the group properties of the space-time
transformations for these accelerated frames differ drastically from those
of the Lorentz group which is defined in the Minkowski space-time
with $\eta_{\mu\nu}=(1,-1,-1,-1)$.  The 
reason is that 
there is simply no equivalence
between inertial and non-inertial frames.  Therefore, the physics of quantum
 fields in these non-inertial frames are much
more involved~\cite{18} and does not have the elegant Lorentz and Poincar\'{e}
invariance.  In view of these differences between inertial and non-inertial
frames, the metric tensor
 $P_{\mu\nu}$ for such a class of general non-inertial frames
may be called the Poincar\'{e} metric tensor.  In the limit of zero
 acceleration, we have $W \to +1$ and $U \to 0$, so that the Poincar\'{e}
metric tensor $P_{\mu\nu}$ in (8) for  non-inertial frames reduces to the
 Minkowski metric tensor $\eta_{\mu\nu}$ for inertial frames.
 
\section{Translation Gauge Symmetry and the Field-Theoretic Origin of 
Effective Metric Tensors}

To see the field-theoretic origin of effective Riemannian metric 
tensors,
 let us consider a fermion field $\psi$.  The kinetic term
  in a fermion Lagrangian in a general frame is given by
$$
i\overline{\psi} \G_\a D^\a \psi - m\overline{\psi} \psi = 
i\overline{\psi} \G_\a \p^\a \psi - m\overline{\psi} \psi,
$$
\be
\{\G_\mu, \G_\nu\} = 2 P_{\mu\nu}(x), \ \ \ \  \G_\mu=\g_a e^a_\mu,
\ee
$$
\{\g_a, \g_b\} = 2 \eta_{ab},  \ \ \ \ \  \eta_{ab} e_\mu^a e_\nu^b =
P_{\mu\nu}.
$$
Here we have the usual relation $D_{\mu}\psi=\p_{\mu} \psi$ because 
the `Lorentz spinor' $\psi$ 
transforms as 
a `coordinate scalar' and $D_\mu$ is the partial covariant derivative defined in terms
of the Poincar\'{e} metric tensor $P_{\mu\nu}$ in a general frame.
 In the presence of the gauge field $\phi_{\mu\nu}$, the translation gauge
symmetry dictates the replacement in a general frame,
$$ D ^\a
\to D ^\a + g\phi^{\a\b} D_{\b} \equiv J^{\a\b}D_\b, \ \ \ \  
J^{\a\b}=P^{\a\b}+g\phi^{\a\b},   $$
\be
 i\overline{\psi} \G_\a D^\a \psi \to  i\overline{\psi} \G_\a \Delta^\a \psi
=  i\overline{\psi} \G_\a J^{\a\b} D_\b \psi =  i\overline{\psi} \g_a
E^{a\b}D_\b \psi
\ee
$$
\Delta^\mu = J^{\mu\a} D_\a, \ \ \ \  E^{a\a} = e^a_\mu J^{\mu\a}.
$$
If one considers $E^{a\a}$ as an `effective tetrad', one has the 
following relation for an `effective metric tensor',
\be
\eta_{ab}E^{a\a} E^{b\b} = \eta_{\a\b} e^a_\mu J^{\mu\a} e^{b\nu} J^{\nu\b}
= P_{\mu\nu} J^{\mu\a} J^{\nu\b} = G^{\a\b}.
\ee
Such an `effective metric tensor' $G^{\a\b}$
also shows up if we consider the Lagrangian of a scalar field $\Phi$ 
with the same replacement as that in (10),
$$\frac{1}{2}[P_{\mu\nu}(D^{\mu}\Phi)(D^{\nu}\Phi) - m^{2}\Phi^{2}] 
\to \frac{1}{2}[G^{\a\b}(D_{\a}\Phi)(D_{\b}\Phi) - m^{2}\Phi^{2}] $$
Therefore, it may appear as if the geometry of the physical space-time
is changed from pseudo-Euclidean space-time to non-Euclidean  space-time
 due to the presence of the
tensor gauge field (or spin-2 field) $\phi_{\mu\nu}$.  However, 
based on the Yang-Mills approach, the presence of $E^{a\a}$ in (10) 
and $G^{\a\b}$ is simply 
the manifestation of the translation gauge
symmetry in flat physical space-time.   

In the literature, when one arrives at this crucial step 
(10),~\cite{7,8,9,10}
one usually gives up the Yang-Mills approach for     a truly   gauge
invariant   theory with a quadratic gauge curvature in a
flat space-time, and follows Einstein's approach to
discuss gravity by postulating Riemannian space-time due to  the presence of
$\phi_{\mu\nu}$ or $J_{\mu\nu}$ in (11).  In other words, one 
postulates $E^{a\a}$ and $G^{\a\b}$ in (11) as a real tetrad and 
a real metric tensor of physical space-time.  Such an approach 
leads to `the most glaring incompatibility of concepts in 
contemporary physics' as observed by Dyson.~\cite{1}
  
We stress that, from 
 the viewpoint of Yang-Mills  theory, the real physical space-time in (10) is still
flat and the fundamental metric tensor is still $P_{\mu\nu}$ in general
frames of reference.  We shall take
this viewpoint throughout the discussion.

All the observable effects of gravity are directly related to the 
 motion of classical objects and light rays.  Thus, it is important 
to understand the relation between the wave equations of fields and the  
corresponding classical equation of a particle and light ray.  In 
Yang-Mills gravity, we show that the equation of motion for a 
classical object is essentially the limit of geometrical optics of 
wave (or field) equations, as demonstrated in the Appendix.  It suffices to say here that if we postulate  the 
following action $S_p$ for classical particles,
\be
S_{p} = - \int mds_{ei}, \ \ \ \ \ \  ds_{ei}^2=I_{\mu\nu}dx^\mu 
dx^\nu, \ \ \ \  I_{\mu\nu}G^{\nu\a}=\delta_{\mu}^{\a},
\ee
$$ G^{\mu\nu} = P_{\a\b} J^{\a\mu} J^{\b\nu}= P^{\mu\nu} +
2g\phi^{\mu\nu} + g^2 \phi^{\mu\ld} \phi^{\nu\s} P_{\ld\s}, 
$$
one can derive the classical equation of motion (i.e., the 
Hamilton-Jacobi equation) which is the same as 
that obtained from 
the classical limit (or the limit of geometric 
optics) of wave equations in the Appendix. 

We consistently treat
 $G^{\mu\nu}$ in (12) as merely an `effective metric tensor' for the
motion of a classical object in flat space-time and in the presence
of  the tensor gauge field.  One can verify that the action $S_p$
(12) is not invariant under the gauge transformation (3).   This is not
surprising because $S_p$ is only an effective action for classical objects
rather than the action for basic tensor and fermion fields.  However,  one
can show that the effective interval $ds^2_{ei}$ and, hence,   the
action $S_p$ are invariant under the following transformation
\be
Q^{\mu_1 ...\mu_m}_{\a_1 ...\a_n} (x) \to Q{^{*}}^{\mu_1 ...\mu_m}_{\a_1
...\a_n} (x)= Q^{\nu_1 ...\nu_m}_{\b_1 ...\b_n} (x)
 \frac{\p x'^{\mu_1}}{\p
x^{\nu_1}}... \frac{\p
x'^{\mu_m}}{\p x^{\nu_m}}
\frac{\p x^{\b_1}}{\p x'^{\a_1}} ... \frac{\p x^{\b_n}}{\p
x'^{\a_n}},
\ee
which corresponds to the transformation (2) with the interpretation (ii),
namely, an arbitrary infinitesimal transformation.  This invariant
property of $S_p$ leads to invariant equation of motion, e.g., the
Hamilton-Jacobi equation, for classical objects  and light rays.  
Furthermore, we
are able to show the agreement between experiments and the action $S_p$ in (12) when the
tensor field $\phi_{\mu\nu}$ is solved from the gauge field equation.  This
agreement will be discussed in sections 6 and 7 below.

\section{Gauge Invariant Action with Fermions, Tensor Fields and 
Quadratic Gauge-Curvature}

Yang-Mills' theory with internal gauge group is 
generalized to a theory with the external gauge group of space-time translation.  In the
generalized Yang-Mills theory with the external translation gauge 
symmetry, we have the gauge curvature
\be
C^{\mu\nu\a} =  J^{\mu\ld}(D_\ld 
J^{\nu\a}) - J^{\nu\ld}(D_\ld J^{\mu\a}),
\ee
which is given by the commutation
relation of the gauge covariant derivative 
$\Delta^\mu=J^{\mu\nu}D_\nu$, $\left[\Delta^\mu, \Delta^\nu \right]=   
 C^{\mu\nu\a} D_{\a}$.

The translation gauge curvature $C_{\mu\a\b}$ involves the symmetric 
tensor gauge field $\phi^{\mu\nu}=\phi^{\nu\mu}$.  Thus, it differs
from the usual Yang-Mills gauge curvature 
$f^k_{\mu\nu} = \p_\nu b^k_\mu - \p_\mu b^k_\nu - b^i_\mu b^j_\nu
c^k_{ij}$, where $c^k_{ij}$ is the structure constant of the gauge group
whose generators have constant matrix representations.  In view of 
this difference, one cannot take any property of the usual Yang-Mills 
theory for granted in the present theory of gravity.  For example, 
although the Yang-Mills theory with internal gauge groups has a 
corresponding fiber bundle, the present Yang-Mills gravity with an external 
space-time group does not.
It can be directly verified that the gauge-curvature $
C^{\mu\nu\a}$ given by (14) satisfies the following identities,
\be
C^{\mu\nu\a}= -C^{\nu\mu\a}, \  \  \  \  \  \   C^{\mu\nu\a} +  
C^{\nu\a\mu} +  C^{\a\mu\nu} = 0,
\ee
because $J^{\mu\nu}=J^{\nu\mu}$.  One can also obtain the Bianchi identity which is complicated for the
translation gauge curvature $C_{\mu\a\b}$.
It turns out that there are only two
independent quadratic gauge-curvature scalars: namely $C_{\mu\a\b}C^{\mu\b\a}$
and $C_{\mu\a}^{ \ \ \  \a}C^{\mu\b}_{ \ \ \  \b}$.  Other
quadratic gauge-curvature scalars can be expressed in terms of them 
because of the identities (15).
We postulate that, in a general frame, the action
$S_{\phi\psi}$ for fermion matter and spin-2 fields involves
the linear combination of the two independent quadratic terms of
the gauge-curvature   and the symmetrized fermion Lagrangian:
\be
S_{\phi\psi}= \int L_{\phi\psi} \sqrt{-P} d^4x, \ \ \ \ \    P=det \ P_{\mu\nu},
\ee
\be
L_{\phi\psi}= \frac{1}{2g^2}\left (C_{\mu\a\b}C^{\mu\b\a}-
 C_{\mu\a}^{ \ \ \  \a}C^{\mu\b}_{ \ \ \  \b} \right) +
\frac{i}{2}[\overline{\psi} \G_\mu \Delta^\mu \psi - (\Delta^\mu
\overline{\psi}) \G_\mu  \psi] -
m\overline{\psi} \psi,
\ee
\be
\Delta^\mu \psi = J^{\mu\nu}D_\nu \psi, \ \ \ \ \  
J^{\mu\nu}=P^{\mu\nu}+
g \phi^{\mu\nu} = J^{\nu\mu} , \ \ \ \ \
D_\ld P_{\mu\nu}=0. 
\ee
Note that the quadratic gauge-curvature term in (17) can also be 
expressed as
\be
L_{\phi\psi}= \frac{1}{2g^2}\left (\frac{1}{2}C_{\mu\a\b}C^{\mu\a\b}-
 C_{\mu\a}^{ \ \ \  \a}C^{\mu\b}_{ \ \ \  \b} \right) +
\frac{i}{2}[\overline{\psi} \G_\mu \Delta^\mu \psi - (\Delta^\mu
\overline{\psi}) \G_\mu  \psi] -
m\overline{\psi} \psi,
\ee
because $
C_{\mu\a\b}C^{\mu\a\b}=2 C_{\mu\a\b}C^{\mu\b\a}.$
The different relative
sign in the two quadratic gauge curvature in the Lagrangian (17) leads to a simple
 linearized equation which is formally  the same as that in general 
 relativity.  
 
Based on the translation   gauge transformation (4), we can shown 
that  
\be
L_{\phi\psi} \to (L_{\phi\psi})^{\$} =L_{\phi\psi} - \L^\ld (\p_\ld L_{\phi\psi}).
\ee
Since $\L^{\mu}$ is an infinitesimal gauge vector function, the gauge 
transformation of $P_{\mu\nu}$ can be written in the form, 
$$
(P_{\mu\nu})^{\$}=P_{\mu\nu}-\L^\ld\p_\ld
P_{\mu\nu}-P_{\mu\b}\p_\nu\L^\b -P_{\a\nu}\p_\mu\L^\a $$
\be
=[(1-\L^\s \p_\s)P_{\a\b}](\delta^\a_\mu -\p_\mu
\L^\a)(\delta_\nu^\b - \p_\nu\L^\b).
\ee
It follows from (21) that
\be
\sqrt{-P} \to \sqrt{-P^{\$}}
= [(1-\L^\s \p_\s)\sqrt{-P}](1 - \p_\ld
\L^\ld), \ \ \ \   P = det P_{\mu\nu}.
\ee
Thus, the Lagrangian $\sqrt{-P}L_{\phi\psi}$  changes only by a divergence under the gauge
transformation,
\be
\int \sqrt{-P}L_{\phi\psi}d^{4}x \to
\int [\sqrt{-P}L_{\phi\psi} -  \p_\ld (\L^\ld L_{\phi\psi} 
\sqrt{-P})]d^{4}x = \int \sqrt{-P}L_{\phi\psi}d^{4}x,
\ee
where we have imposed a constraint $D_\mu \L^\mu=(1/\sqrt{-P})\p_\mu (\sqrt{-P} \L^\mu)=0$ and 
the volume element $ \sqrt{-P}d^{4}x$ is invariant. The divergence term in (23) does not contribute to field equations because
one can transform an integral over a
4-dimensional volume into the integral of a
vector over a hypersurface on the boundaries of the volume of integration
 where fields and their variations vanish.
Thus, we have shown that the action $S_{\phi\psi}$ in (16) is invariant 
under the $T_4$ gauge transformations (4).  This result holds in both inertial and non-inertial frames.

\section{ Tensor and Fermion Equations in General Frames}

In general, field equations with gauge symmetry are not well defined.
 One usually includes a 
suitable gauge-fixing
term in the Lagrangian to make the solutions of gauge field equation 
well-defined.  Yang-Mills gravity
in a general frame is based on the  total
Lagrangian $L_{tot}\sqrt{-P}$, which is the original Lagrangian with an additional
 gauge-fixing term $L_{gf} \sqrt{-P}$ involving  the gauge parameter  $\xi$:
\be
L_{tot}\sqrt{-P} = \left( L_{\phi\psi} + L_{gf} \right)\sqrt{-P},
\ee
\be
L_{gf}=\frac{\xi}{2g^2}[D_\mu J^{\mu\a} - \frac{1}{2}D^\a J^\mu_\mu][D^\nu 
J_{\nu\a} - \frac{1}{2}D_\a J^\nu_\nu].
\ee
Here, the gauge-fixing term corresponds to the usual gauge condition 
$\p_{\mu} \phi^{\mu\nu} - (1/2)\p^{\nu} \phi^{\ld}_{\ld} = 0$ for tensor 
fields.  The Lagrange equations for the gravitational tensor field 
$\phi^{\mu\nu}$ in a general frame can be derived from the action
$\int L_{tot} \sqrt{-P} d^4x$, we obtain
\be
H^{\mu\nu} + \xi A^{\mu\nu}=  g^2 T^{\mu\nu},
\ee
$$
H^{\mu\nu} \equiv D_\ld (J^{\ld}_\rho C^{\rho\mu\nu} - J^\ld_\a 
C^{\a\b}_{ \ \ \ \b}P^{\mu\nu} + C^{\mu\b}_{ \ \ \ \b} J^{\nu\ld})  
$$
\be
- C^{\mu\a\b}D^\nu J_{\a\b} + C^{\mu\b}_{ \ \ \ \b} D^\nu J^\a_\a -
 C^{\ld \b}_{ \ \ \ \b}D^\nu J^\mu _\ld,
\ee
\be
A^{\mu\nu} = D^\mu \left(D^\ld J_\ld{^\nu}  - \frac{1}{2} D^\nu 
J^\ld_\ld \right)  - \frac{1}{2} P^{\mu\nu} \left(D^\a D^\ld J_{\ld\a}  
 - \frac{1}{2} D^\a D_\a J^\ld_\ld \right), 
\ee
where $\mu$ and $\nu$ should be made symmetric.   
We have used the identity (15) in the derivation of (26).  The energy-momentum tensor $T^{\mu\nu}$ of fermion matter and the partial covariant
derivative $D_\ld$  associated with the Poincar\'{e} metric tensor
$P_{\mu\nu}$ are given by
\be
T^{\mu\nu} = \frac{1}{2}\left[ \overline{\psi} i\Gamma^\mu D^\nu \psi -
i(D^\nu \overline{\psi}) \Gamma^\mu \psi \right], \ \ \ \  D^\nu \psi =
\p^\nu \psi,
\ee
\be
D_\ld J^{\mu\nu} = \p_\ld J^{\mu\nu} + \Gamma^\mu_{\ld\rho}
J^{\rho\nu} +  \Gamma^\nu_{\ld\rho} J^{\mu\rho}, \ \ etc.
\ee
where the Christoffel symbol is given by $
\Gamma^\mu_{\ld\rho} = \frac{1}{2} P^{\mu\s}(\p_\ld P_{\s\rho} +   
\p_\rho P_{\s\ld} - \p_\s P_{\ld\rho}).$

The Dirac 
equation for a fermion interacting with the tensor fields 
$\phi^{\mu\nu}$ in a general frame  can also be derived from (24):  
\be
\begin{split}
& i\G_\mu (P^{\mu\nu}+g\phi^{\mu\nu})D_{\nu} \psi - m \psi + \frac{i}{2} [D 
_{\nu}(J^{\mu\nu} \G_{\mu})]\psi = 0,\\ 
& i(P^{\mu\nu}+g\phi^{\mu\nu})(D_{\nu} \overline{\psi}) \G_{\mu} + m 
\overline{\psi} + \frac{i}{2} \overline{\psi}[D 
_{\nu}(J^{\mu\nu} \G_{\mu})] = 0, 
\end{split}
\ee
where we have used the relation   
$(1/\sqrt{-p})\p_{\nu}[Q^{\nu}\sqrt{-P}]=D_{\nu}Q^{\nu}$.
If one compares the fermion equation (31) with the 
 Dirac equation in quantum electrodynamics [i.e., $ (i\g^\mu \p _\mu - e 
 \g^\mu A_\mu - m )\psi = 0$] in inertial frames, one can see a distinct difference:  Namely, 
 the kinematic term $i\g^\mu \p_\mu$ and the electromagnetic coupling 
 term $e\g^\mu A_\mu$ have a different relative sign, if one takes  the 
 complex conjugate of the Dirac equations.  This implies the presence of 
 both repulsive and attractive forces between two charges.
  However, if one takes  the complex conjugate of the fermion 
  equation for $\psi$ in (31), there is no change in 
 the relative sign of the kinematical term and the spin-2 coupling 
 term. Thus, the translation gauge symmetry of gravity naturally
 explains the universal attractive force of gravity for 
 fermion matter and anti-fermion matter.  

In inertial frames with $P_{\mu\nu} = \eta_{\mu\nu}$,
the gauge-field equation (27) can be linearized as follows:
\be
\p_\ld \p^\ld \phi^{\mu\nu} -  \p^\mu \p_\ld \phi^{\ld\nu} -
\eta^{\mu\nu} \p_\ld \p^\ld \phi  + \eta^{\mu\nu} \p_\a \p_\b \phi^{\a\b}
+  \p^\mu \p^\nu \phi - \p^\nu \p_\ld \phi^{\ld\mu} - g T^{\mu\nu} = 0,
\ee
for weak fields.  This equation can also be written in the form:
\be
\p_\ld \p^\ld \phi^{\mu\nu} - \p^\mu \p_\ld \phi^{\ld\nu} +
 \p^\mu \p^\nu \phi^\ld_\ld - \p^\nu \p_\ld \phi^{\ld\mu} = g (T^{\mu\nu}
- \frac{1}{2}\eta^{\mu\nu}
 T^\ld_\ld),
\ee
where we have set $\xi=0$ and used
$J^{\mu\nu} = \eta^{\mu\nu} + g \phi^{\mu\nu}.$  It is interesting to 
see that the
linearized gauge-field equation (32) is formally the same as the
corresponding equation in general relativity.  This property may be 
related to the fact that the transformation (2) is formally the same as that in 
general relativity.

\section{ The Perihelion Shift with a New Correction Term}

In order to show possible differences between Yang-Mills gravity and general 
relativity, let us consider the perihelion shift to the second order 
approximation.  The perihelion shift can be seen from the solution of 
the Hamilton-Jacobi equation for a classical particle.  The momentum $p_{\mu}$ of a classical particle can 
be derived from the action $S_{p}$ (12), we have
\be
p_{\nu}= -\frac{\p S_{p}}{\p x^{\nu}} = -m\frac{dx^{\mu}}{ds_{ei}} 
I_{\mu\nu},
\ee
Since $ds_{ei}^2=I_{\mu\nu}dx^\mu dx^\nu,$ we have
\be
G^{\mu\nu}p_{\mu} p_{\nu} - m^{2} = 0, \ \ \ \ \   I^{\mu\nu}
G_{\nu\ld} = \delta^\mu_\ld.
\ee
The result (34) is obtained on the basis of the variation of the 
particle action (12)
\be
\delta S_p = - m I_{\mu\nu}
\left(\frac{dx^\mu}{ds_{ei}}\right)
\delta x^\nu,
\ee
which can be derived if we consider only  the actual path with one of its end point
variable.  From equations (34) and (35), we obtain
the following Hamilton-Jacobi equation for a particle with mass m,
\be
G^{\mu\nu}(\p_\mu S)(\p_\nu S) - m^2 = 0,  \ \ \ \  S \equiv S_{p}.
\ee
This equation can also be obtained from the Dirac equation 
in the presence of the gravitational tensor 
field $\phi^{\mu\nu}$ by considering the classical 
limit, which resembles the limit of geometric optics.
  (See Appendix.)  
Equations (35) and (37) are formally  the same
 as the corresponding equations in general relativity.

Equation (32) or (33) with the only non-vanishing component,
$T_{00}=m\delta^3 ({\bf r})$, leads to $g\phi^{00} = g^2 m/(8\pi r)$ 
in the Newtonian limit.  Also,
$G^{00}$ in the Hamilton-Jacobi equation (37) should have the usual result $G^{00} = 1 +
2G m/r$ in this limit, where $G$ is the gravitational constant.
Based on these results, together with $G^{\mu\nu}$
in (37) and $I_{\mu\nu}$ in (12), we obtain the first order
 approximation in an inertial frame,
\be
g=\sqrt{8\pi G}, \ \ \ \ \  and  \ \ \ \ \  g\phi_{00} = g\phi_{11}
= \frac{G m}{r}, \ \ \ \
etc.
\ee
These results can be obtained by solving (33) with the spherical coordinate, $x^\mu=(w,
r, \theta, \phi)$.

Let us consider
 the perihelion shift of Mercury, which is sensitive to the
coefficient appearing in the second-order term of $G^{00}$ or 
$I_{00}$ and in the first-order term of $G^{11}, G^{22}$ and $G^{33}$.  However, we shall calculate  the second-order terms of 
all  components for the effective metric tensors $G^{\mu\nu}$ to show 
that the observable result is gauge invariant, i.e., independent of 
the gauge parameter $\xi$.  We
solve the non-linear gauge field equations by the method of successive approximation
and carry out the related post-Newtonian approximation
to a second order.  For gauge field equations
to be well defined to the second order, it is
convenient to use  gauge-field equation (26) with the gauge parameter 
$\xi$. 

For simplicity, we consider an inertial frame and a static and spherically symmetric system, in
which  tensor gauge fields are produced by a spherical object at rest  with mass $m$. 
Based on symmetry considerations,~\cite{23} the 
non-vanishing components of the exterior solutions $\phi^{\mu\nu}(r)$ are 
$\phi^{00}(r), \phi^{11}(r), \phi^{22}(r)$ and 
$\phi^{33}(r)=\phi^{22}/ 
sin^2 \theta $, where $x^\mu =(w,r,\theta, \phi)$. To solve the static 
gauge field, let us write
$
J^{00}= J^0_0  = S, \   -J^{11}= J^1_1= R,$ and   $- r^{2}¥J^{22}= 
J^2_2= - r^2 \sin^2\theta J_{33}=J^3_3 = T.
$
The metric tensor is given by $P_{\mu\nu} = (1,   -1, -r^2, 
-r^2 \sin^2\theta )$.  In this coordinate system, the
 non-vanishing components of the Christoffel symbol 
$\G^\a_{\mu\nu}$ are given by $
\G^1_{22} = -r, \   \G^1_{33} = -r  \sin^2\theta, \   
\G^2_{12}  = 1/r, \ 
\G^2_{33} = -\sin\theta \ \cos\theta,  \   \G^3_{13}  = 1/r, 
\   \G^3_{23} = \cot\theta.$

After some tedious but straightforward calculations, the gauge field 
equation (26) with $(\mu,\nu) = (0,0), 
(1,1), (2,2), (3,3)$, can be written respectively as
$$
\frac{d}{dr}\left(R^2 \frac{dS}{dr}\right) + \frac{2}{r} R^2 \frac{dS}{dr} + 
\left(R\frac{d}{dr} + \frac{dR}{dr} + \frac{2}{r} R\right)\left(-R(\frac{dS}{dr} + 
2\frac{dT}{dr}) - \frac{2}{r} T^2 + \frac{2}{r} TR\right) 
$$ 
\be
+ \ \  \xi\left[ \frac{1}{4} \frac{d^2 }{dr^2}(S - R +2T)  + 
\frac{1}{2r}\frac{d}{dr} (S - 3R+4T) - \frac{1}{r^2} (R-T) 
\right ]  = 0, 
\ee
$$
R(\frac{dS}{dr})^2 + 2r^3\frac{d(T/r)}{dr}\left[\frac{R}{r}\frac{d(T/r)}{dr} + 
\frac{T^2}{r^3}\right] + \left(\frac{dS}{dr} + 2\frac{dT}{dr} - \frac{2T}{r}\right)\times
$$
$$\left(-R(\frac{dS}{dr} + 
2\frac{dT}{dr}) - \frac{2}{r} T^2 + \frac{2}{r} TR \right) 
$$
\be
+ \ \  \xi \left[ \frac{1}{4} \frac{d^2 }{dr^2}(S - R +2T)  - 
\frac{1}{2r}\frac{d}{dr} (S +R) + \frac{3}{r^2} (R-T)  
\right] = 0, 
\ee
$$
\left(R\frac{d}{dr} + \frac{dR}{dr} + \frac{5R}{r} - \frac{2T}{r}\right)\left[\frac{R}{r}\frac{d(T/r)}{dr} + 
\frac{T^2}{r^3}\right] 
$$
$$
+\left[\frac{1}{r^2}(R\frac{d}{dr} + 
\frac{dR}{dr}) + \frac{3R}{r^3} - \frac{2T}{r^3} \right] \left[-R\left(\frac{dS}{dr} + 
2\frac{dT}{dr}\right) - \frac{2}{r} T^2 + \frac{2}{r} TR\right] 
$$
\be
+ \ \ \xi \left[\frac{1}{4r^2} \frac{d^2 }{dr^2}(S - R +2T)  - 
\frac{1}{r^3}\frac{d}{dr} (R-T) + \frac{1}{r^4} (R-T)  \right] = 0. 
\ee
 The equation for
$(\mu,\nu) = (3,3)$ is the same as that in (41).  

We can solve the gauge field equations (39)-(41) to a second order
approximation by setting $S=1+a_{o}/r+a/r^{2}$, $R=1+b_{o}/r + 
b/r^{2}$, etc.   We obtain the second-order approximation of
the tensor field which satisfies  the gauge field equation (26),
$$
g \phi^{00} = \frac{G m}{r} + \frac{G^2 m^2}{2r^2}, \ \ \ \ \
g \phi^{11} = \frac{G m}{r} + \frac{K_{1}}{r^2}, $$  
\be
g \phi^{22} = - \frac{1}{r^2} \left[-\frac{G m}{r} + \frac{K_{2}}{ 
r^2}\right],  \ \ \ \ \  g \phi^{33} = g \phi^{22}/sin^2 \theta;
\ee
$$   K_{1}= 
\left(\frac{2}{\xi} + \frac{1}{2} \right)G^2 m^2, \ \ \ \   
K_{2}=2G^2 m^2 \left(\frac{1}{\xi} - 1 \right),  $$
Note that the first order approximation is independent of the gauge
parameter $\xi$.  
  We have seen that only the second order terms in $\phi^{11}, 
  \phi^{22}$ and $\phi^{33}$ depend
on the gauge parameter $\xi$.  However, all the first order
terms and the second order term in $\phi^{00}$ do not depend on the gauge
parameter $\xi$, and these are the only crucial terms for the
observable results of the perihelion shift.

From the result (42) and $G^{\mu\nu}(r)$ given in (12) with
 $P_{\mu\nu}  = (1, -1, -r^2, -r^2 sin^2 \theta)$, we obtain the effective 
 metric tensor,
$$
G^{00}(r)= 1 + \frac{2G m}{r}
+ \frac{2G{^2} m^2}{r^2},  \ \ \ \ \   G^{11}(r)=-\left[1 - \frac{2G m}{r} + 
 \frac{L_{1}}{r^2}  \right],    
$$
\be
G^{22}(r)=- \frac{1}{r^2} \left(1 - \frac{2G m}{r} + 
 \frac{L_{2}}{r^2} \right), \ \ \ \ \ \ \  G^{33}(r)= G^{22}(r)/sin^2 \theta;
\ee
$$ L_{1} =  - \frac{4}{\xi} G^2 m^2, \ \ \    L_{2} = \left(\frac{4}{\xi} - 
    3\right)G^2 m^2.$$
These results for effective metric tensors are well defined in the 
limit $\xi \to \infty$.  This particular gauge may be called `static 
gravity gauge.'  If one chooses the static gravity gauge, the 
effective metric tensors are given by
\be
G^{00}(r)= 1 + \frac{2G m}{r}
+ \frac{2G{^2} m^2}{r^2},  \ \ \ \ \   G^{11}(r)=-\left[1 - \frac{2G m}{r}  
  \right], 
\ee
$$
G^{22}(r)=- \frac{1}{r^2} \left(1 - \frac{2G m}{r} - 
 \frac{3G^{2}m^{2}}{r^2} \right), \ \ \ \ \ \ \  G^{33}(r)= G^{22}(r)/sin^2 \theta.
$$

Let us carry out   the calculation of the perihelion shift   to the second 
order for all components of $G^{\mu\nu}(r)$ in 
(43) in terms of the spherical coordinates $x^\mu = (w, \rho, \theta, 
\phi)$.  This can be accomplished by a change of variable
$ \rho^2 = r^2/(1 - 2Gm/r + L_2/r^2) = - G^{22}(r)$, 
where $G^{22}(r)$ is given in (43). We obtain
\be
r= \rho B, \ \ \ \ \ \ \ \ \ \   B \equiv \left[1-\frac{Gm}{\rho} + 
\frac{G^{2}m^2}{2\rho^2}(\frac{4}{\xi} -6)\right],   
\ee
$$
dr = 
d\rho\left[1-\frac{G^{2}m^{2}}{2\rho^{2}}(\frac{4}{\xi} -6)\right].
$$
The effective metric tensor $G^{\mu\nu}(\rho)$ in the Hamilton-Jacobi 
equation (37) (with $r=\rho$) is obtained in the spherical coordinate
$x^\mu = (w, \rho, \theta, \phi)$ as follows:
$$
G^{00}(\rho)=G^{00}(r)|_{r=\rho B}= 1 + \frac{2Gm}{\rho}
+ \frac{4G^{2}m^{2}}{\rho^2},  $$
\be
G^{11}(\rho)=G^{11}(r)¥(\frac{d\rho}{dr})^{2}¥|_{r=\rho 
B}¥ = - \left[1 - \frac{2Gm}{\rho} - 
 \frac{8G^{2}¥m^{2}¥}{\rho^2} \right], 
\ee
$$G_{22}(\rho)=- \rho^2,  \ \ \ \ \ \  G_{33}(\rho)= -\rho^2 sin^2 \theta.
$$
Note that   the gauge parameter $\xi$ in $G^{11}(r)¥$ and $(d\rho /dr)^{2}¥$  
cancel each other  
so that $G^{11}(\rho)¥$ is $\xi$-independent, in agreement with gauge invariance. 
Therefore,  all components of $G^{\mu\nu}(\rho)$ for the spherical 
coordinates are independent 
of the gauge parameter $\xi$ to the second-order approximation.  As 
far as experiment is concerned, the result (46) is effectively 
equivalent to that of general relativity.~\cite{23}  The second=order 
term in $G^{11}(\rho)$ differs from that in general relativity and 
leads to a slightly different prediction for the perihelion shift, as 
we shall see below.

To see the physical implications of (46), we choose $\theta$ = $\pi/2$ so that the 
Hamilton-Jacobi equation (37)  for a 
planet  with mass $m_p$ has the following form:
\be
G^{00}(\rho)\left(\frac{\p S}{\p w} \right)^2 + G^{11}(\rho)\left(\frac{\p S}{\p 
\rho}\right)^2 + G^{33} (\rho) \left(\frac{\p S}{\p \phi}\right)^2 - m_p^2 = 0.
\ee
According to  the general procedure  for solving the 
Hamiltonian-Jacobi equation, we write the solution of S 
 in the form $S= -E_o w + M \phi + f(\rho)$.~\cite{24}  We solve for 
 $f(\rho)$, and obtain
\be
S  =  - E_ow + M\phi + \int \frac{1}{\sqrt{|G^{11}|(\rho)}} \sqrt{E_o^2 
G^{00}(\rho) -   m_p^2 - \frac{M^2}{\rho^2}} d\rho,
\ee
where $E_o$ and $M$ are respectively  constant energy and angular 
momentum of the planet.
  The trajectory is determined by $\p S/\p M =  
constant$, so that we have
\be
\phi = \int \frac{(M/\rho^2) d \rho}{\sqrt{E_o^2 
G^{00}|G^{11}| -   m_p^2 
|G^{11}| - M^2 |G^{11}|/\rho^2}}. 
\ee
To find   the trajectory, it is convenient to write (49) as a differential 
equation of $\sigma = 1/\rho $.  
We obtain      the following equation
\be
\frac{d^2 \sigma}{d\phi^2} = \frac{1}{P} - \sigma (1+ Q) + 3Gm\sigma^2,
\ee
$$ P= \frac{M^2}{m_p^2 Gm}, \ \ \ \ \  Q=  \frac{8Gm}{P} \left( 
\frac{E_o^2 - m_p^2}{m_p^2} \right),$$
by differentiating the equation with respect to $\phi$.  Thus we see 
that the equation for the trajectory (50) in Yang-mills gravity 
differs slightly from the corresponding equation in general 
relativity by  a new correction term Q.
This correction term $Q$ is of the order of $(Gm/P)\b^2$ which 
is   undetectable because of      the velocity $\b$ of 
the planet is very small in comparison with the speed of light, $\b 
<< 1.$

By   the usual successive approximation,~\cite{24} we obtain  the solution
\be
\sigma = \frac{1}{P(1+Q)} \left[ 1+ e \ cos\left(\phi  
(1 - \frac{3Gm}{P} + \frac{Q}{2} \right) \right]. 
\ee
The advance of the perihelion for one revolution of the planet is  give by
\be
 \delta \phi = \frac{6\pi Gm}{P} \left(1 - 
\frac{3(E_o^2 - m_p^2)}{4 m_p^2} \right) , 
\ee
We note that    the second term in the bracket of (52) shows    the difference between  the present 
Yang-Mills gravity and Einstein's theory.  This result shows that    the 
observable perihelion shift is independent of the gauge parameter $\xi$ 
which appears in the second order approximation of the solution of 
$g\phi^{\mu\nu}$.  Since   the observational accuracy of the perihelion shift of 
the Mercury is about 1 $\%$,  the prediction (52) of Yang-Mills gravity  can be 
tested only if the Mercury were to move with a tenth of the speed of 
light such that $(E_o^2 - m_p^2)/m_p^2 \approx \b^2 \approx 0.01 $. 
 It is highly unlikely for a macroscopic planet to have such a speed.  Thus, the result 
(52) of Yang-Mills gravity is consistent with existing data for the 
perihelion shift.~\cite{25}

\section{ Bending of Light and Other Experiments}

The bending of light can be derived from the propagation of
a light ray in geometrical optics in an inertial frame.  Suppose the light
ray propagates in the
presence of the tensor gauge fields,  its path is determined by the
eikonal equation, 
\be
G^{\mu\nu} \p_\mu \Psi \p_\nu \Psi = 0.  
\ee
This eikonal equation can  be directly 
derived from the Maxwell's equation in the presence of the 
gravitational tensor field $\phi^{\mu\nu}$ in the limit of 
geometrical optics.  (See Appendix.) 
It can also be obtained from  the Hamilton-Jacobi equation 
(37) with $m \to 0$ and $\p_\mu S \to \p_\mu \Psi$, where $\Psi$ 
is the eikonal.  As usual, we assume   that  the motion of the light ray
is in a plane passing through the origin and having the angle $\theta =
\pi/2$.  Using $G^{\mu\nu}(\rho)$ given in (46) and $x^\mu = (w, \rho, \theta, 
\phi)$, the eikonal
 equation (53) can be written as
\be
I^{00}\left(\frac{\p \Psi}{\p w} \right)^2 + I^{11}\left(\frac{\p \Psi}
{\p \rho}\right)^2 - \frac{1}{\rho^2} \left(\frac{\p \Psi}{\p \phi}\right)^2 = 0.
\ee
By the general procedure of solving (54) in a spherical symmetric tensorÊ
field, we look for the eikonal $\Psi$ in the form~\cite{24}
\be
\Psi = -E_0 w + M\phi + f(\rho).
\ee
 One can determine $f(\rho)$ and solve  for the trajectory of
the ray, which is the same as 
(50) with
$m_p \to 0$ and $E_0$ replaced by $\omega_{o} = - \p \Psi/\p w$ 
(c=1).  We have
\be
\frac{d^2 \sigma}{d\phi^2} = - \sigma (1+ Q_{o}) + 3Gm\sigma^2, \ \ 
\  \sigma=\frac{1}{\rho}
\ee
\be
Q_{o}=  \frac{8G^{2}m^{2}}{M^{2}} ,
\ee
where  the  new correction term $Q_{o}$  
is extremely small. 
Following the usual procedure~\cite{24}, we find the following result 
for the deflection of a light ray,
\be
\Delta \phi \approx \frac{4Gm\omega_o}{M} \left(1 - 
\frac{18 G^{2}m^{2} 
\omega_o^{2}}{M^{2}} \right),
\ee
We note that the additional correction term in the bracket differs from 
that in general relativity and is 
negligible for the bending of light by the Sun.  
A ray of light passing through  a spherical symmetry
tensor field at a distance $R_{o}$ from the center of the sun will have a
deflection $\Delta \phi \approx 4Gm/R_{o} \approx 1.75^{\prime \prime}$,
to the first order approximation.  This
result is consistent with experiment and is also the same as that obtained
in general relativity, as one would expect based on    the results (46)
 for the effective metric tensor.
 
 Historically, red shift and time dilatation due to the gravitational 
 effect were
originally derived  by using the principles of 
equivalence.~\cite{26}Ê
Nevertheless, these experiments can also be discussed 
within the present framework of Yang-Mills gravity in flat space-time:  Both    the
red shift and the time dilatation caused by gravity (or the tensor
gauge  fields) can be considered as  physical results of the invariance of the
effective action (12) or the `proper time' $\tau = \int ds_{ei}$ under the
transformation (13), without assuming
the usual principle of equivalence.Ê

With the help of geometrical optics,~\cite{24}   the red shift
can also be derived from  the  eikonal equation (53).
This eikonal equation is fundamental in geometrical optics and is
invariant under the transformation (13).  For static
tensor field, $G^{\mu\nu}$ does not contain time $w=x^0$, so that the
frequency $k^c_0 = -\p \Psi/\p w$ is constant during the propagation of the
light ray.~\cite{24}  On the other hand,  the frequency $k_0 =  -
\p \Psi/\p \tau$ measured in terms of the `proper time' depends on
positions in space.  Thus, we have
\be
k_0 = - \frac{\p \Psi}{\p w}\frac{\p w}{\p \tau}=Ê
\frac{k^c_0}{\sqrt{I_{00}}}, \ \ \ \   I_{00}=\frac{1}{G^{00}} \approx 1 - 2g\phi^{00}.
\ee
In fact, this relation with $G^{00}$ given in 
(46) for the spherical coordinate is the same as that 
in general relativity and is consistent with the experiment of 
red-shift.~\cite{23,24}

The experiment of the time delay of radar echoes passing the sun 
can be explained by the gauge field
equation (26) under the simplifying assumption of isotropy and time
independence.  Specifically, this experiment can be explained by the result 
$G^{11}(\rho)$ in (46) to the first order in
$Gm/\rho$.~\cite{23}Ê
The effective metric tensor in (46) is the same as that obtained in general
relativity to the first order approximation.  Thus, if one follows the usual procedure
of calculations,~\cite{23} one can verify that Yang-Mills gravity is 
also consistent with the experiment of radar echoes.

In Yang-Mills gravity, the gravitational quadrupole radiations of 
binary pulsars can be 
calculated to the second-order in $g\phi^{\mu\nu}$ in inertial frames.  
The energy-momentum tensor of gravitation $t_{\mu\nu}$ is defined by the 
field equation (26) (with $\xi = 0$) written in the following form,
$D^\ld D_{\ld} 
\phi^{\mu\nu}= -g(T^{\mu\nu} + t^{\mu\nu}), $
 in a general frame. Using the usual approximations   and gauge  condition 
 $\p_{\mu} \phi^{\mu\nu}= \p^{\nu} \phi^{\ld}_{\ld}/2$, we can calculate 
the average energy-momentum of a gravitational plane wave and the
 power by the usual method.~\cite{23}  For example,
the power $P_o$ emitted per unit solid angle in the direction ${\bf x}/|{\bf x}|$ 
 can be written as 
\be
\frac{dP_o}{d\Omega}= \frac{G \omega^2}{\pi}\left(T^{\ld\rho}({\bf k},\omega)
 T^*_{\ld\rho}({\bf k},\omega) -
 \frac{1}{2}T({\bf k},\omega)T^{*}({\bf k},\omega) \right), 
 \ee
where $T({\bf k}, \omega)$ is defined as follows:~\cite{23}
Suppose one observes    this radiation in the wave zone, one can write  
the polarization tensor in terms of the Fourier transform of $T_{\mu\nu}$:
\be
e_{\mu\nu}({\bf x},\omega)=\frac{-g}{4\pi r}[T_{\mu\nu}({\bf k},\omega) - 
\frac{1}{2} \eta_{\mu\nu}T ({\bf k},\omega)], \ \ \ \  T= T^\ld_\ld, 
\ee
\be
T_{\mu\nu}({\bf k},\omega) \equiv \int d^3{\bf x}' T_{\mu\nu}({\bf 
x}',\omega)]exp(-i{\bf k} \cdot{\bf x}'), 
\ee
where the polarization tensor $e_{\mu\nu}({\bf x},\omega)$ is defined by 
the relation: 
\be
\phi_{\mu\nu} ({\bf 
x,t}) \approx [e_{\mu\nu}({\bf x}, \omega) exp(-ik_\ld x^\ld) + c.c.]. 
\ee
To the second-order approximation, the result (60) for 
the power emitted per solid angle in Yang-Mills 
gravity turns out to be the same as that 
obtained in general relativity and consistent with the data of the 
binary pulsar PSR 1913+16.~\cite{23,27,28} 

\section{Remarks and Discussions}

Although the invariant action of Yang-Mills gravity is dictated by the 
space-time translation gauge symmetry, the gauge-fixing 
term in the action is not.  We observe that the relation between the 
effective metric tensor 
$G^{\a\b}$ in Hamilton-Jacobi equation and the tensor field 
$\phi^{\mu\nu}$ appears to be dependent on the 
choice of the specific form of the gauge-fixing term.  For example, 
suppose one chooses
\be
L_{gf}\sqrt{-P} = \left( \frac{\eta}{2g^2}(D_\mu J^{\mu\a})D^\nu
J_{\nu\a}\right)\sqrt{-P},
\ee
where $\eta$ is a gauge parameter.  The gauge-dependent terms in the 
static field equation (40) 
will be modified as follows:
\be
R(\frac{dS}{dr})^2 + 2r^3\frac{d(T/r)}{dr}\left[\frac{R}{r}\frac{d(T/r)}{dr} +
\frac{T^2}{r^3}\right] + \left(\frac{dS}{dr} + 2\frac{dT}{dr} - \frac{2T}{r}\right)\times
\ee
$$\left(-R(\frac{dS}{dr} +
2\frac{dT}{dr}) - \frac{2}{r} T^2 + \frac{2}{r} TR \right) -
 \eta
\frac{d}{dr}(\frac{dR}{dr} + \frac{2R}{r}  - \frac{2T}{r}) = 0,
$$
Naturally, the solution will be different from those given in (42).  
Thus, if one wants to preserve  the result (46),  the relation 
between the effective metric tensor $G^{\a\b}$ and the tensor field 
$\phi^{\mu\nu}$ has to be modified 
accordingly.  This property 
suggests that  the types of gauge-fixing terms that can be used for 
external space-time translation symmetry are more restricted than those in 
the Yang-Mills theory with internal gauge groups. 

So far, there is no observable difference between Yang-Mills 
gravity and Einstein's theory in known experiments within the solar 
system, it is possible that the difference between the two theories 
can be tested by observations of phenomena outside the solar system.  For example, 
the binary pulsar PSR 1913+16 provides an interesting and unique test of 
gravitational theories.~\cite{25,27,28}     Both the pulsar and its silent companion 
are about 1.4 times the mass of the Sun.  They travel with a speed 
that range up to $4\times 10^{5}$ meters per second in a tight orbit 
with a minimum separation roughly equal to the radius of the Sun. 
As a result, the binary 
pulsar has a very large advancing of periastron, 4.2 degrees per 
year.~\cite{29}  We have examine this case and we find that 
the data is not accurate 
enough to test the difference between Yang-Mills gravity and 
general relativity.   The quadrupole radiation and experiments 
related to the binary pulsar 
will be discussed in detail in a separate paper.Ê

The theory of Yang-Mills gravity in flat space-time
 has a well-defined conservation law for the energy-momentum tensor.
 The space-time translation symmetry
plays an essential role in connecting the tensor Yang-Mills field to its
source, i.e., the conserved energy-momentum tensor (through the Noether
theorem).  Furthermore, it is gratifying that the Hamilton-Jacobi 
equation (37)
 for a classical particle can also be derived from the corresponding 
 fermion wave equation, as shown in Appendix.  Thus, Yang-Mills 
gravity in flat space-time reveals a more well-defined theory and
a more coherent relation between its quantum and classical 
 aspects than conventional formalisms in curved space-time. 
 
In previous attempts to formulate a gauge theory of gravity, one usuallyÊ
followed Einstein's approach based on Riemannian space-time and obtained
Einstein's equation or closely related field equations for the metric
tensor.~\cite{5,6,7,8,9,10}  As a result, the gauge 
symmetry, however powerful it may be,
was unable to simplify the complicated interaction terms and, hence, the resultant
theory of gravity had serious ultraviolet divergences and was not renormalizable.  In
view of this difficulty,
 we follow closely the Yang-Mills approach with a quadraticÊ
gauge-curvature and formulate the theory of Yang-Mills gravity on  the basis of
 the translation gauge symmetry and flat space-time.
 
\section{Conclusions}

The Yang-Mills approach to gravity reveals an interesting 
property.  Namely, the 
action (12) for the motion of a classical particle with the effective metric
 $ds_{ei}^2=I_{\mu\nu}dx^\mu dx^\nu$ (or the Hamilton-Jacobi equation 
 (A.7) with $G^{\mu\nu}$ in the Appendix) 
 shows that the
underlying basis for gravity is probably the translation gauge 
symmetry in a flat space-time rather than theÊ
general coordinate invariance in curved space-time.   

In general, the basic Lagrangians for vector and tensor  fields in 
Yang-Mills gravity do 
not explicitly and unambiguously involve the effective metric tensor $G^{\mu\nu}$. 
Yang-Mills gravity 
reveals the field-theoretic origin of an `effective Riemannian metric tensor'
only in the limit of geometrical optics or classical limit of the 
wave equations, as shown in (A.3) and (A.7) in the Appendix.  
Therefore, the effective metric tensor $G^{\mu\nu}$ does not play any 
basic role in the quantum aspect of Yang-Mills gravity. 

Based on previous discussions, we conclude that Yang-Mills gravity is viable because
the gravitational gauge equation (26) is consistent with all known experiments of
 gravity.

\bigskip

\noindent
{\bf Acknowledgements}
\bigskip

The work is supported in part by the Jing Shin
Research Fund of the UMass Dartmouth Foundation.  He would like to thank
Leonardo Hsu and colleagues at the National Center for Theoretical
Sciences (NCTS, Taiwan), National Taiwan University, and UMass Dartmouth for
their discussions.

\bigskip

\noindent
  
{\bf Appendix. \ \    Derivations of Eikonal 
 and Hamiltonian-Jacobi Equations}
\noindent
\bigskip

In Yang-Mills gravity, the fundamental equation (53) of geometrical
optics can be derived as follows:  We postulate the translation gauge
invariant Lagrangian $L_{em}$ for the electromagnetic potential $A^{\mu}$,
$$\hspace{0.4in}
L_{em}=-\frac{1}{4}P^{\mu\a}P^{\nu\b}F_{\mu\nu}F_{\a\b}, \ \ \
\ \  F_{\mu\nu}=\Delta_\mu A_{\nu}-\Delta_\nu A_{\mu}, \ \ \ \ \
\Delta_\mu=J_{\mu\nu}D^{\nu}, \hspace{0.4in}  (A.1)$$
where we have used the same replacement in (10).  For
simplicity, let us consider an inertial frame with
$P^{\mu\nu}=\eta^{\mu\nu}$ and $\Delta_{\mu}=J_{\mu\ld}\p^{\ld}$,
one can obtain the wave equation,
$$ \hspace{0.9in} 
\Delta_\mu(\Delta^{\mu} A^{\ld}-\Delta^{\ld}A^{\mu}) +
(\p_{\a}J^{\a}_{\mu})(\Delta^{\mu} A^{\ld}-\Delta^{\ld}A^{\mu})=0.
 \hspace{0.9in} (A.2)$$
Using the electromagnetic gauge condition, $\p_{\mu}
A^{\mu}=0,$ and the expression for the field
$A^{\ld}=a^{\ld}exp(i\Psi)$, we can derive the eikonal equation
(53),
$$ \hspace{1.1in}
G^{\mu\nu} \p_\mu \Psi \p_\nu \Psi = 0, \ \ \ \ \ \ \  Ê
G_{\mu\nu} = P_{\a\b} J^{\a\mu} J^{\b\nu}, \hspace{1.2in} (A.3)$$
in the limit of geometrical optics.  That is, both the eikonal
$\Psi$ and the wave 4-vector $\p_{\mu}\Psi$ are very 
large.~\cite{30}  We stress that the Lagrangian (A.1) and  the wave 
equation (A.2) do not imply an effective
metric tensor $G^{\mu\nu}$.  Only in 
 the limit of geometrical optics of the wave equation (A.2), an effective metric tensor 
 $G^{\mu\nu}$ emerges. 

Next, let us consider the relation between the Hamilton-Jacobi 
equation (37) and the massive fermion wave equation.  The fermion 
wave equation (31) can be derived from the Lagrangian (24), i.e.,
$$\hspace{1.1in}
i\G_\mu \Delta^\mu \psi - m \psi + \frac{i}{2} \g_{a}[D 
_{\nu}(J^{\mu\nu} e^{a}_{\mu})] \psi = 0. \ \ \  \hspace{1.3in} (A.4) $$
Using the expression for the field
$\psi= \psi_{o}exp(iS)$, we can derive the equation
$$ \hspace{1.2in}
\g_{a} E^{a\mu} \p_{\mu} S + m  - \frac{i}{2} \g_{a}[D 
_{\nu}(J^{\mu\nu} e^{a}_{\mu})] = 0. \ \ \  \hspace{1.2in}  (A.5) $$

In the classical limit, the momentum $\p_{\mu} S$ and mass $m$ are 
large quantities, and one can neglect the small gravitational interacting 
term involving $e^{a}_{\mu}$.
To eliminate the spin variables, we multiply a 
factor $(\g_{a} E^{a\mu} \p_{\mu} S - m)$ to the large terms in 
 (A.5), the resultant equation can be written in the form
$$\hspace{1.0in}  
\frac{1}{2}(\g_{b}\g_{a} + \g_{a}\g_{b})E^{a\mu} 
E^{b\nu}(\p_{\mu}S)(\p_{\nu}S) - m^{2} = 0. \hspace{1.1in} (A.6)$$
With the help of the anti-commutation relation for $\g_{a}$ in (9) and the 
effective metric tensor (11), (A.6) leads to the Hamilton-Jacobi 
equation,
$$ \hspace{1.6in} G^{\mu\nu}(\p_{\mu}S)(\p_{\nu}S) - m^{2} = 0, 
\hspace{1.7in} (A.7)$$
for the motion of a classical particle in the presence of the 
gravitational tensor field $\phi^{\mu\nu}$.  It is important  that 
the result (A.7) is consistent with 
equation (37) obtained from the particle action $S_{p}$ in 
(12).\footnote{This property can be treated properly by a similar 
fashion in a previous discussion in Ref. 31.}

\bigskip


\bibliographystyle{unsrt}

\end{document}